\documentclass[letter, 12pt]{article}

\usepackage[margin=1in]{geometry}
\usepackage[numbers,sort&compress]{natbib} 
\usepackage{authblk} 
\usepackage{color}
\usepackage{amsmath}
\usepackage{amssymb} 
\usepackage{graphicx} 
\usepackage{fancyhdr} 
\usepackage{amsfonts}
\usepackage{tabularx} 
\usepackage[
singlelinecheck=false 
]{caption}

\def\be{\begin{equation}}
\def\ee{\end{equation}}
\def\bea{\begin{eqnarray}}
\def\eea{\end{eqnarray}}
\def\nn{\nonumber}

\def\ba{\begin{array}}
\def\ea{\end{array}}

\begin{document}

\rhead{OHSTPY-HEP-T-15-005}
\renewcommand\headrule{} 

\title{\huge \bf{Gluino LOSP with Axino LSP}}
\author{Stuart Raby}
\affil{\emph{Department of Physics}\\\emph{The Ohio State University}\\\emph{191 W.~Woodruff Ave, Columbus, OH 43210, USA}}

\maketitle
\thispagestyle{fancy}
\pagenumbering{gobble} 

\begin{abstract}\normalsize\parindent 0pt\parskip 5pt
In this letter we have presented a novel version of ``long-lived" gluinos in supersymmetric models with the gluino the lightest ordinary supersymmetric particle [LOSP] and axino LSP.   Within certain ranges of the axion decay constant $f_a < 1 \times 10^{10}$ GeV, the gluino mass bounds are reduced to less than 1000 GeV.   The best limits can be obtained by looking for decaying R-hadrons in the detector where the gluino decays to a gluon and axino in the calorimeters.   SUSY models with a gluino LOSP can occur over a significant region of parameter space in either {\em mirage-mediation} or general gauge-mediated SUSY breaking models.  The gluino LOSP is not constrained by cosmology, but in this scenario the axion/axino may be good dark matter candidates.

\end{abstract}

\pagenumbering{arabic} 

\newpage
\section{Introduction}

Supersymmetric models with non-universal gaugino masses are well-defined even in the range of parameters
for which gluinos are the LSP.  The most stringent constraint in this case comes from cosmology.   Although gluinos are Majorana particles and annihilate with
a strong interaction rate,  there are still too many of them left over in the early universe.  They form stable,  hadronic
bound states and searches for heavy hydrogen essentially rule out this possibility \cite{Muller:1976qy,Smith:1979rz,Smith:1982qu}.

There are presently strong limits on gluino masses coming from the first run of the LHC.  In particular, we are interested in
the bounds for long-lived gluinos.  We consider recent CMS results.  These come from searches for R-hadrons which travel through the detector \cite{Chatrchyan:2013oca} or for R-hadrons which stop in the detector and then decay \cite{Khachatryan:2015jha}.
For long-lived gluinos the bounds depend on the fraction, $f$, which initially hadronize as a gluino-gluon bound state \cite{Chatrchyan:2013oca}.
For $f = 0.1$, the bounds require $m_{\tilde g} > 1233$ GeV or even greater, depending on the model for propagation.   These bounds only require data from the central tracker.  The track that the charged R-hadron leaves in the central tracker must satisfy $|\eta| < 2.1$, $p_T > 45$ GeV/c and have significant $dE/dx$.  The track must also be isolated in both the tracker and calorimeter with $\sum p_T < 50$ GeV/c where the sum is over all
tracks (except the candidate track) within a cone $\Delta R < 0.3$ radians.  Thus if the R-hadron should decay in the tracker or calorimeter it will
not pass these cuts. However, if the lifetime is long enough such that the gluino decays outside the window of the collision region, it will be counted in this event sample.  Therefore, in order to evade these bounds, the gluino must decay fast enough so that it leaves energy in the calorimeters. Thus we need $\langle \beta \rangle \ c \tau_{\tilde g} \lesssim 2.5$ meters or $\tau_{\tilde g} \lesssim 8.3 \ (\langle \beta \rangle)^{-1} \times 10^{-9}$s.   If the gluino does live long enough to make it to the muon system, then it will be constrained by other data sets which give just as strong bounds on the gluino mass.  The limits are slightly less strong for gluinos which are long-lived and stop in the detector \cite{Khachatryan:2015jha}.  In this case the most stringent bound is $m_{\tilde g} \geq 1000$ GeV for gluino life-times in the range, $10^{-6} s < \tau_{\tilde g} < 10^{3} s$.  However,  any event which satisfies the latter stopped bounds were already constrained by the former long-lived bounds.

These limits can be applied to any theory where the gluino is the NLSP which decays into a gluon and neutral LSP.   It has applications to gauge-mediated SUSY breaking models where the gluino decays to a gluon and a goldstino \cite{Dimopoulos:1996vz} or to split SUSY models where the gluino decays to a quark-anti-quark pair and neutralino LSP \cite{ArkaniHamed:2004fb}.

In this letter we consider another scenario where the gluino is the lightest ordinary supersymmetric particle [LOSP] with
an axino LSP.    The strong CP problem has a natural solution in terms of the Peccei-Quinn-Weinberg-Wilczek axion \cite{Peccei:1977hh,Weinberg:1977ma,Wilczek:1977pj}.  In a supersymmetric theory the axion has a scalar partner, the saxion, and
a fermionic partner, the axino.   The saxion typically has mass of order the gravitino mass while the axino can be much lighter.

Supersymmetric models with non-universal gaugino masses, such as {\em mirage-mediation} SUSY breaking or general gauge-mediation
have significant ranges of parameters where the gluino is the LSP.  For example, in a recent study of {\em mirage-mediation} in
an SO(10) GUT \cite{Anandakrishnan:2014nea} the gaugino mass formula at the GUT scale is given by
\be
M_i = \left( 1 + \frac{g^2_G b_i
\alpha}{16\pi^2}\text{log}\left(\frac{M_{Pl}}{m_{16}}\right) \right) M_{1/2} \; ,
\ee
where $b_i=(33/5,1,-3)$ for $i=1,2,3$, $M_{1/2}$ is the overall mass scale, and $\alpha$ is the ratio of the anomaly mediation to gravity mediation contributions.  For $\alpha$ in the range, $3 \leq \alpha \leq 4$, we find that the gluino is the LSP.   Also in the case of general gauge-mediation (see Ref. \cite{Raby:1998xr}),  for messenger scales of order the GUT scale,  the gravitino is heavy and we find that it is quite natural to have a gluino LSP.
We note that these models have the very interesting property of precision gauge coupling unification \cite{Raby:2009sf,Krippendorf:2013dqa}, i.e. the gauge couplings unify at the GUT scale to high accuracy.   Thus requiring little or no threshold corrections at $M_{GUT}$.
However, as discussed earlier, such theories are unacceptable cosmologically.   So therefore it becomes advantageous to combine these models with
the axion solution to the strong CP problem.  Then the gluino is the LOSP with an axino LSP.

\section{Gluino - Axino coupling and gluino life-time}

We consider the supersymmetrized axion multiplet coupling to the SU(3) gauge sector.  We have \cite{Strumia:2010aa,Choi:2011yf}
\be  {\cal L}^{eff} =  - \frac{\alpha_s}{2 \sqrt{2} \pi f_a} \int d^2\theta \ A \ W^{a \alpha} \ W^a_\alpha + h.c. \ee
where the axion supefield,  \be A = \frac{s + i a}{\sqrt{2}} + \sqrt{2} (\theta \ \psi_a) + (\theta)^2 \ F_A , \ee
and the gauge superfield strength is given by \be W^a_\alpha = - i \lambda^a_\alpha + [\delta^\beta_\alpha \ D^a - \frac{i}{2} (\sigma^\mu \bar \sigma^\nu)^\beta_\alpha G_{\mu \nu}^a ] \theta_\beta + (\theta)^2 \ \sigma^\mu_{\alpha \dot{\alpha}} \ D^\mu \ \bar \lambda^{\dot{\alpha}} . \ee

We define the four component Majorana spinors for the gluino and axino fields by
\be \tilde g = \left( \ba{c} - i \lambda \\ i \bar \lambda \ea \right),  \;\;\; \tilde a =   \left( \ba{c}  \psi_a \\ \bar \psi_a \ea \right) . \ee
With this notation, we have
\bea {\cal L}^{eff} = & \frac{\alpha_s}{8 \pi f_a} [ a \ (G^{a \mu \nu} \ \tilde G^a_{\mu \nu} + D_\mu (\bar \tilde g \ \gamma^\mu \gamma_5 \ \tilde g) & \\ & + s \ (G^{a \mu \nu} \ G^a_{\mu \nu} - 2 D^a \ D^a + 2 i \bar \tilde g \ \gamma^\mu \ D_\mu \ \tilde g) & \nn \\ & +
i \bar \tilde a \ G_{\mu \nu}^a \frac{[\gamma^\mu, \gamma^\nu]}{2} \ \gamma_5 \ \tilde g - 2 \bar \tilde a \ \tilde g \ D^a ] & \nn \\ &
+ \sqrt{2} F_A \ (\lambda \ \lambda) + h.c. .&  \nn \eea

We then find the gluino decay rate (neglecting the axino mass) given by
\be \Gamma_{\tilde g \rightarrow \tilde a \ g} = \frac{\alpha_s^2 \ m_{\tilde g}^3}{128 \pi^3 \ f_a^2} . \ee
Thus the gluino lifetime for particular values of $f_a$ and gluino mass is given by
\be \tau _{\tilde g \rightarrow \tilde a \ g} = 3 \times 10^{-8} s \frac{(f_a/10^{10} \; {\rm GeV})^2}{(m_{\tilde g}/850 \; {\rm GeV})^3} . \ee
Therefore, as an example, for values of $f_a \leq 0.7 \times 10^{10}$ GeV and an average gluino velocity, $\beta \sim \frac{1}{2}$,  the gluino mass bound may reduce to $m_{\tilde g} \gtrsim 850$ Gev.

\section{Conclusion}

We have presented a novel version of gluinos LOSPs in supersymmetric models with an axino LSP.   Within certain ranges of the axion decay constant $f_a < 1 \times 10^{10}$ GeV, the gluino mass bounds are reduced to less than 1000 GeV.   The best limits can be obtained by looking for decaying R-hadrons in
the hadronic calorimeters.  Such decays will produce jets deep in the calorimeter.  SUSY models with a gluino LOSP can occur over a significant region of parameter space in either {\em mirage-mediation} or general gauge-mediate SUSY breaking models.  Since the gluino LOSP is no longer constrained by cosmology,  it would now be interesting to analyze the possibility of axion/axino dark matter in this scenario.

\section*{Acknowledgments}
S.R. acknowledges stimulating conversations with L.J. Hall,  F. D'Eramo and significant feedback from C.S. Hill during the preparation of this letter.
S.R.~received partial support for this work from DOE/ DE-SC0011726.   The idea was first presented while S.R. was supported by the Munich Institute for Astro- and Particle Physics (MIAPP) of the DFG cluster of excellence "Origin and Structure of the Universe".

\clearpage
\newpage

\newpage
\appendix

\clearpage
\newpage

\bibliographystyle{utphys}
\bibliography{bibliography}

\end{document}